\documentstyle[12pt,twoside]{article}
\def\bA{{\bar{\alpha}}}
\def\bB{{\bar{\beta}}}
\def\bG{{\bar{\gamma}}}
\def\bE{{\bar{\epsilon}}}
\newcommand{\norm}[1]{{\protect\normalsize{#1}}}
\newcommand{\LAP}
{{\small E}\norm{N}{\large S}{\Large L}{\large A}\norm{P}{\small P}}
\newcommand{\ben}{\begin{equation}}
\newcommand{\een}{\end{equation}}
\newcommand{\bea}{\begin{eqnarray}}
\newcommand{\eea}{\end{eqnarray}}
\newcommand{\nn}{\nonumber \\ }
\newcommand{\hf}{\frac{1}{2}}
\newcommand{\eq}{\begin{equation}}
\newcommand{\en}{\end{equation}}
\newcommand{\eqn}{\begin{eqnarray}}
\newcommand{\enn}{\end{eqnarray}}

\newcommand{\g}{{\bf g}}

\newcommand{\E}{\eta}

\newcommand{\pa}{\partial}

\newcommand{\toz}{\frac{\T_{12}}{Z_{12}}}
\newcommand{\tr}{{\rm tr}}

\newcommand{\A}{\alpha}
\newcommand{\B}{\beta}
\newcommand{\D}{\delta}

\newcommand{\G}{\gamma}
\newcommand{\EP}{\epsilon}

\newcommand{\r}{{\cal R}}
\newcommand{\Si}{\sigma}
\newcommand{\T}{\theta}

\newcommand{\NPB}[1]{{\it Nucl. Phys.} {\bf B#1}}
\newcommand{\PLB}[1]{{\it Phys. Lett.} {\bf B#1}}

\newcommand{\IJMPA}[1]{{\it Int. J. Mod. Phys.} {\bf A#1}}
\newcommand{\CMP}[1]{{\it Comm. Math. Phys.} {\bf #1}}

\def\hJ{\hat{J}}
\def\hM{\hat{M}}

\newcommand{\C}{\mbox{\hspace{1.24mm}\rule{0.2mm}{2.5mm}\hspace{-2.7mm} C}}

\newcommand{\Z}{\mbox{$Z\hspace{-2mm}Z$}}

\newcommand{\und}[1]{\underline{#1}}
\newcommand{\mb}[1]{\hs{5}\mbox{#1}\hs{5}}
\newcommand{\hs}[1]{\hspace{#1 mm}}
\begin{document}
\begin{titlepage}
\null

\vspace{7mm}
\begin{center}
  {\Large\bf Quantum Hamiltonian Reduction \par}
  {\Large\bf in Superspace Formalism \par}
  \vspace{1.5cm}
  \baselineskip=7mm

  {\large Jens Ole Madsen\footnote{\noindent email : jomadsen@nbivax.nbi.dk \\
Address after March 1, 1994 :
Niels Bohr Inst., Blegdamsvej 17, DK-2100 Copenhagen {\O}, Denmark.}
and
Eric Ragoucy\footnote{\noindent email : ragoucy@lapvax.in2p3.fr} \par}

\vspace{5mm}

{\sl Laboratoire de Physique Th\'eorique \LAP, \\
groupe d'Annecy, LAPP, Chemin de Bellevue, B.P. 110, \\
F-74941 Annecy-le-vieux Cedex, France. \par}

\vspace{3cm}

{\bf Abstract}
\end{center}
\par

Recently the quantum hamiltonian reduction was done in the case of general
$s\ell(2)$ embeddings into Lie algebras and superalgebras. In this paper we
extend the results to the quantum hamiltonian reduction of $N=1$ affine Lie
superalgebras in the superspace formalism. We show that if we choose a gauge
for the supersymmetry, and consider only certain equivalence classes of fields,
then our quantum hamiltonian reduction reduces to quantum hamiltonian
reduction of non-supersymmetric Lie superalgebras. We construct explicitly the
super energy-momentum tensor, as well as all generators of spin 1 (and
$\hf$); thus we construct explicitly all generators in the superconformal,
quasi-superconformal and $\Z_2 \times \Z_2$ superconformal algebras.

\begin{flushright}
\LAP-A-459/94 \\
February 1994
\end{flushright}

\end{titlepage}
\setcounter{footnote}{0}

\renewenvironment{thebibliography}[1]
  { \begin{list}{\arabic{enumi}.}
    {\usecounter{enumi} \setlength{\parsep}{0pt}
     \setlength{\itemsep}{3pt} \settowidth{\labelwidth}{#1.}
     \sloppy
    }}{\end{list}}

\section{Introduction.}

\indent

It is well known, that for every $s\ell(2)$ embedding into a Lie
(super)algebra, one can find a $W$-algebra by hamiltonian reduction of the
corresponding WZNW model, see e.g. \cite{BeOo,BaTjVDr,BaFeFoRaWi,BoSc}.
The quantization of
the $W$-algebras obtained from principal $s\ell(2)$ embeddings
has been studied in
a number of papers, e.g. \cite{FeFr,BoMcPi,FrKaWa}, and for arbitrary
embeddings
of $s\ell(2)$ into Lie algebras
the quantization of these $W$ algebras has recently been performed by de Boer
and Tjin \cite{BoTj}, using the BRST formalism as well as the theory of
spectral sequences. For general
$s\ell(2)$ embeddings into Lie superalgebras the problem is slightly more
complicated, because of the necessity in many cases to introduce auxiliary
fields, but this problem was solved in \cite{SeThTr,SeTr}.

In \cite{DeRaSo,FrRaSo} it was shown that if we wish to consider supersymmetric
$W$ algebras in the $N=1$ superspace formalism, then we must consider instead
$osp(1|2)$ embeddings into Lie superalgebras. The purpose of this paper is to
apply the methods used in \cite{BoTj} to the supersymmetric case, thereby
quantizing the $N=1$ $W$ algebras, giving at the same time the so-called
quantum Miura transformation that produces free-field realizations of these
algebras.
A description of the $N=1$ superspace formalism that we use can be found in e.g
\cite{DiKnPeRo}.

In section \ref{s.2} we show the supersymmetric quantum
hamiltonian reduction, using a
super BRST operator. In section \ref{s.3} we show that if we choose
a gauge for the
supersymmetry, and consider not all the fields in the operator product algebra
but only certain equivalence classes of fields, then the super BRST operator
reduces to the BRST operator found in \cite{SeThTr,SeTr} in the case of the
quantum hamiltonian reduction of non-supersymmetric Lie superalgebras.
In section \ref{s.4} we construct explicitly the super energy
momentum tensor, as well
as all the spin 1 generators of any $W$-algebra, and thus we construct all the
generators in the superconformal, quasi-superconformal and $\Z_2 \times \Z_2$
superconformal algebras \cite{FrLi,ItMaPe,FrRaSo}, and we comment on the
quantum Miura transformation.
In section \ref{s.5} we consider in more detail the superconformal algebra
obtained
from the hamiltonian reduction of $B(n|1) = osp(2n+1|2)$, and in section
\ref{s.6} we
sum up our results.

\section{Supersymmetric BRST Cohomology.\label{s.2}}

\indent

In the case that we are considering, we have a basic classical simple Lie
superalgebra $\g = \g_0 \oplus \g_1$, $\g_0$ is the bosonic subalgebra and
$\g_1$ is the fermionic subspace. We
denote the generators of this Lie superalgebra by $t^a$.
The supersymmetric, invariant bilinear form on $\g$ we write as
$g^{ab} = \langle t^a, t^b \rangle$, and the structure constants are
$[t^a,t^b] = {F^{ab}}_c t^c$ ($[\cdot,\cdot]$ is the graded commutator of the
Lie superalgebra); here and in the following summation over repeated indices
is implied. For $t^a \in \g_\A, \A \in \Z_2$ we define
$d_a = \A+1$.

We define the associated affine Lie algebra
$\g^{(1)}$ with level $k$ in terms of the operator product expansions:
\ben
J^a(Z) J^b(W) = \frac{k g^{ab}}{Z-W} + \frac{(\T-\E)}{Z-W}{f^{ab}}_c\ J^c(W) +
\cdots,
\een
where the dots stands for the regular part.
${f^{ab}}_c$ are defined by ${f^{ab}}_c = (-)^{(d_a+1)d_b} {F^{ab}}_c$
\cite{DeRaSo}, and the superspace variables are $Z=(z,\T)$, $W=(w,\eta)$.
Note that the $d_a$'s defined above are the usual $\Z_2$ gradings for the
currents $J^a$: $J^a(Z)J^b(W) = (-)^{d_ad_b} J^b(W) J^a(Z)$.

We use the tensor
$g^{ab}$ and its inverse $g_{ab}$ to raise and lower
indices, e.g. $f^{abc} = {f^{ab}}_{c'}g^{c'c}$. We notice that we have
$f^{abc} \not = 0 \Rightarrow d_a + d_b + d_c = 1$ (mod 2)   and
$g^{ab} \not = 0 \Rightarrow d_a = d_b$.

One can show the following symmetry properties of the structure constants and
the bilinear form:
\bea
{f^{ab}}_c & = & (-)^{d_ad_b} {f^{ba}}_c           \nn
g^{ab} & = & (-)^{d_a+1} g^{ba}              \nn
f^{abc} = {f^{ab}}_e g^{ec} & = & (-)^{d_bd_c} f^{acb}.
\eea

Now let us consider an embedding of $osp(1|2)$ into the Lie superalgebra $\g$,
and let us denote the generators corresponding to this embedding by
$t^i$, $i \in \{--,-,0,+,++\}$.
Under this embedding $\g$ splits up into representations $\r_j$ of $osp(1|2)$,
$j$ being the spin of the representation. This gives a natural grading of the
generators, namely the eigenvalue
under the adjoint action of $t^0$ (sometimes called the magnetic quantum
number $m$). It will be
convenient in most of the paper to assume that we have chosen a basis for $\g$
such that all the generators belong to one of the representations $\r_j$.
Let us define
$\Delta_m$ to be the set of indices of generators with grade $m$:
$a \in \Delta_m \Leftrightarrow [t^0,t^a] = m~t^a$.

We can write the current of the affine Lie superalgebra in the form
$J(Z) = \sum (-)^{d_a+1} J^a(Z) t_a$. The constraints that we wish to impose
are \cite{DeRaSo,FrRaSo} $J^-(Z) = \chi^-$, $J^a(Z) = \chi^a = 0$ for $a$ with
negative
grade, $a \not = -$, where $\chi^-$ is some constant. For simplicity we will
take $\chi^- = 1$. We will use the convention
that we write indices with negative grade with greek letters, and indices
with non-negative grade with barred greek letters. We can then write the
constrained current as
\bea
J(Z) & = & \sum_\A (-)^{d_\A+1} \chi^\A t_\A
+ \sum_{\bA} (-)^{d_\bA+1} J^{\bA}(Z) t_{\bA} \nn
& = & -t_- + \sum_{\bA} (-)^{d_\bA+1} J^{\bA}(Z) t_{\bA}
\eea

We now wish to quantize this constrained system. There are in principle two
methods for doing this. We can impose the constraints on the classical affine
Lie superalgebra, and then
quantize the resulting system; or we can use the BRST formalism to impose the
constraints on the quantized affine Lie superalgebra. The disadvantage of the
first method is that in general the constrained system is quite complicated,
and even though it can be quantized in some cases, there is no systematic
way of
doing it in the general case. On the other hand, we know very well how to
quantize any affine Lie superalgebra, and there is a general method for using
the BRST formalism to impose such constraints: the constrained quantum algebra
is the zeroth cohomology of a certain BRST operator \cite{KoSt,FeFr}.

{}For each constraint $\chi^\A$ we introduce a
pair of superghosts $(B^\A,C_\A)$
of conformal dimensions $(\hf,0)$, with Grassmann gradings $(d_\A+1,d_\A)$ and
ghost numbers $(-1,1)$. We define $F(\Omega)$ to be the operator
product algebra generated by the affine currents $J^a$ and the ghosts $B^\A$
and $C_\A$, and we note that $F(\Omega)$ is graded by the ghost number.
The BRST operator that we need is then given by:
\bea
S_{BRST} O(W) & = & \oint_W dZ J_{BRST}(Z) O(W),\nn
J_{BRST}(Z) & = & (J^\A(Z) - \chi^\A) C_\A(Z)
+ \hf (-)^{d_\A+d_\B} {f^{\A\B}}_{\G} (B^\G C_{\B} C_{\A})_0(Z).
\eea

This is a straightforward generalization of the
BRST-current given in \cite{BoTj}. $(\cdot)_0$ denotes normal ordering.
Note that
$J_{BRST}$ is a bosonic current, which implies that $S_{BRST}$ is a
fermionic operator.

$S$ has ghost number one, and as mentioned above we wish to calculate the
zeroth cohomology
$H^0(F(\Omega),S_{BRST})$.
The BRST transformations are explicitly given by the following expressions :
\bea
S_{BRST} J^a & = & -kg^{a\B}DC_\B + {f^{a\B}}_c J^c C_\B \nn
S_{BRST} C_\A & = & \hf (-)^{(d_\B+d_\G)} {f^{\B\G}}_\A C_\G C_\B \nn
S_{BRST} B^\A & = &
(-)^{d_\A} \left ( J^\A + (-) ^{d_\A + d_\B} {f^{\A\B}}_\G B^\G C_\B
\right ) - \chi^\A
\label{brst-trans}
\eea
Straightforward calculations show that indeed these transformations are
nilpotent.

Following \cite{BoTj} we now split the BRST operator into two :
$S_{BRST} = S_0 + S_1$,
where $J_0(Z) = - \chi^\A C_\A(Z)$ and
$J_1(Z) = J^\A(Z)C_\A(Z)
+ \hf (-)^{d_\A+d_\B} {f^{\A\B}}_{\G} ( B^\G C_{\B} C_{\A})_0(Z)$. If we
define $\hJ^a =  J^a + (-) ^{d_a + d_\B} {f^{a\B}}_\G B^\G C_\B$, we can
write the transformations $S_0$ and $S_1$ in the form
\bea
S_0 \hJ^a & = & - (-)^{d_a+d_\B} {f^{a\B}}_\G \chi^\G C_\B \quad = \quad
{f^{a\B}}_{-} C_\B \nn
S_0 C_\A & = & 0 \nn
S_0 B^\A & = & - \chi^\A
\label{brst-trans0}
\eea
and
\bea
S_1 \hJ^a & = & - k g^{a\B} DC_\B + {f^{a\B}}_{\bar{\G}} \hJ^{\bar{\G}}C_\B \nn
S_1 C_\A & = & \hf (-)^{d_\B+d_\G} {f^{\B\G}}_\A C_\G C_\B \nn
S_1 B^\A & = & (-)^{d_\A}    \hJ^\A.
\label{brst-trans1}
\eea
The ``hatted" generators obey the same OPE's as the unhatted ones; this is
in contrast
to the non-supersymmetric case, where the addition of the ghost-part changes
the level.
{}From the transformations above, we can show that
$S_0^2 = S_0S_1 = S_1S_0 = S_1^2 = 0$. We can now define a bi-grading of the
generators of the algebra :
\ben
\left.
\begin{array}{rcl}
\mbox{deg}(J^a)  & = & (2m,-2m) \\
\mbox{deg}(C_a) & = & (-2m,2m+1) \\
\mbox{deg}(B^a) & = & (2m,-2m-1)
\end{array}
\right \} \mbox{  for }a \in \Delta_m.
\een
With these definitions $D_0$ has grade $(1,0)$, and $D_1$ has grade $(0,1)$,
and with this bigrading $F(\Omega)$ becomes a double complex. In the case of
$s\ell(2)$
embeddings into Lie algebras, the cohomology of this complex was found in
\cite{BoTj}. We find that the supersymmetric case is very similar, so we will
only give a very brief description of
the formalism. For the details we refer the reader to \cite{BoTj}.

As we see from eq. (\ref{brst-trans}) we have
$S_{BRST} B^{\A} = (-)^{d_\A} \hJ^\A - \chi^\A$ and
$S_{BRST} \hJ^\A = 0$. In the case of Lie algebras, this
implies that $F^\A(\Omega)$ generated by $\hJ^\A$ and
$b^\A$ is a subcomplex of $F(\Omega)$.
In the case of Lie superalgebras a slight complication
arises, because we may have a fermionic root $\A$ such that $2\A$ is also a
root. In that case we have
$\hJ^\A(Z) \hJ^\A(W) = \frac{\T-\E}{Z-W}{f^{\A~\A}}_{2\A} \hJ^{2\A}(W)$ and
$\hJ^\A(Z) B^\A(W) = \frac{\T-\E}{Z-W}{f^{\A~\A}}_{2\A} B^{2\A}$.
This means that $F^{\A}(\Omega)$ is not a subcomplex, and we must consider
instead $F^{\A,2\A}(\Omega)$ generated by
$\hJ^\A$, $\hJ^{2\A}$, $B^\A$ and $B^{2\A}$. One still finds, however, that the
cohomology of this subcomplex is trivial:
$H^q(F^{\A,2\A}(\Omega);S_{BRST}) = \C\delta_{q,0}$.

We now define $F_{red}(\Omega)$ to be the algebra generated by the remainder
of the generators, i.e. by $\hJ^{\bA}$ and $C^\A$. One can then show
that
\bea
H^*(F(\Omega);S_{BRST}) & \simeq &
H^*(F_{red}(\Omega);S_{BRST}) \otimes
\left ( \bigotimes
H^*(F^{\A,2\A}(\Omega);S_{BRST}) \right ) \nn
& & \otimes
\left ( \bigotimes H^*(F^{\A}(\Omega);S_{BRST}) \right ) \nn
& = & H^*(F_{red}(\Omega);S_{BRST}).
\eea

The next step is to find the cohomology of $F_{red}(\Omega)$. We find that
$S_0$ annihilates $J^\bA$ only if $t^\bA$ is a highest weight in an $osp(1|2)$
representation, and that $C_\A$ is $S_0$-trivial for all $\A$; thus we have
\ben
H^q (F_{red}(\Omega);S_0) \simeq F_{hw}(\Omega)\D_{q,0}.
\een
where $F_{hw}(\Omega)$ is the subspace generated by highest weight generators
(generators that have regular operator product expansions with $J^+$).
Using the theory of spectral sequences, one can show \cite{BoTj} that this
implies
$$
H^q(F_{red}(\Omega);S_{BRST}) \simeq F_{hw}(\Omega)\D_{q,0}.
$$
This gives us the
final result for the cohomology:
\ben
H^q(F(\Omega), S_{BRST}) \simeq F_{hw}(\Omega)\D_{q,0}.
\een
In order to find representatives for the
elements in the cohomology, i.e. generators of the $W$ algebra, one uses the
so-called tic-tac-toe construction. One starts an element of
$F_{hw}(\Omega)$,
a highest weight generator of grade $j$: $J_{hw}^\bA$, $\bA \in \Delta_j$.
{}From
each such generator one can construct a
generator of the $W$ algebra by
$$
W^\bA(Z) = \sum_{i=0}^{2j} (-)^i W^\bA_i (Z),
$$
where $W^\bA_0 = J_{hw}^\bA$ and $S_1 W^\bA_i = S_0 W^\bA_{i+1}$.
It is clear that
the series stop as indicated at $W^\bA_{2j}$, since this element has grade zero
which implies that $S_1 W^\bA_{2j} = 0$;
in fact one can show \cite{BoTj} that the term $W^\bA_{2j}$ is non vanishing
for all generators $W^\bA$ in the $W$-algebra.
It is easy to check
that this is really an element in $H^0(F(\Omega), S_{BRST})$
and it is clear that
these elements form a basis of $H^0(F(\Omega), S_{BRST})$.

\section{``Cohomological" Factorization of Spin $\hf$ Fields.\label{s.3}}

\indent

When we consider the classical hamiltonian reduction defined by an $osp(1|2)$
embedding into a Lie superalgebra $\g$,
one of the authors has
shown \cite{Ra} that if we choose a suitable gauge for the supersymmetry and
factorize the spin 1/2 fields in the algebra (see also \cite{DeTh}), the result
is the $W$-algebra one gets by performing the hamiltonian reduction defined
by the embedding of $s\ell(2) \subset osp(1|2)$ into $\g$ in the
non-supersymmetric case.
A similar result can be shown in the quantum hamiltonian reduction.
The comparison of the two approaches is done using a filtration
of the cohomology of $S_{BRST}$. As we will break the $N=1$
supersymmetry, we need to expand superfields into fields.
\ben
\begin{array}{ll}
\widehat{J}^a(z,\theta)= \widehat{\psi}^a(z) +\theta\
\widehat{\jmath}^a(z)
& J_{BRST}(z,\theta)= j_0(z) +\theta\  j_1(z) \\
C_\alpha(z,\theta)= c_\alpha(z) +\theta\  \gamma_\alpha(z)
& B^\alpha(z,\theta)= \beta^\alpha(z) +\theta\  b^\alpha(z)
\end{array}
\een
We also introduce $\Pi_0$, the subset of $\Delta_0$ corresponding to
$\r_0$ representations:
\ben
a\in\Pi_0\ \ \Leftrightarrow\ \ a\in\Delta_0 \mb{and} [t^+,t^a]=0
\een
and decompose the set of barred greek indices into three pieces:
$\Delta_+$, $\Pi_0$, and $(\Delta_0-\Pi_0)$ with the notation
$\und{\alpha}$, $i$ and $\bar{\imath}$ for the three sets respectively.

At the field level, the cohomology of $S_{BRST}$ (with current
$J_{BRST}$) becomes the cohomology of $s$ with current $j_1$.
Explicitly, we have
\ben
\begin{array}{c}
s(\widehat{\psi}^a)=-k g^{a\beta}\, \gamma_\beta +{f^{a\beta}}_{\bar{\gamma}}
\, \widehat{\psi}^{\bar{\gamma}}\, c_\beta - (-)^{d_a+d_\beta}\,
{f^{a\beta}}_{\gamma}\, \chi^\gamma\, c_\beta  \\
s(\widehat{\jmath}^a)= kg^{a\beta} \partial c_\beta
-{f^{a\beta}}_{\bar{\alpha}} \left( \widehat{\jmath}^{\bar{\alpha}}\,
c_\beta +(-)^{d_{\bar{\alpha}}}\,  \widehat{\psi}^{\bar{\alpha}}\,
\gamma_\beta\right) +(-)^{d_a+d_\beta}\, {f^{a\beta}}_\alpha\, \chi^\alpha
\, \gamma_\beta \\
s(\beta^\alpha)= (-)^{d_\alpha}\ \widehat{\psi}^\alpha -\chi^\alpha
\hfill s(b^\alpha)= (-)^{d_\alpha+1}\ \widehat{\jmath}^\alpha\\
s(\gamma_\alpha)= (-)^{d_\beta+d_\mu+1}\ {f^{\beta\mu}}_\alpha\,
\gamma_\mu\, c_\beta
\hfill  s(c_\alpha)= \frac{1}{2} (-)^{d_\beta+d_\gamma}\
{f^{\beta\gamma}}_\alpha\, c_\gamma c_\beta
\end{array}
\een
Now, we introduce a filtration of the cohomology of $s$ on
$F(\Omega)=F_0$, i.e. a sequence of subcomplexes $F_i$
of $F(\Omega)$ such that $F_0\supset F_1\supset F_2\supset
\dots$  Here, the filtration we consider is almost trivial, since it
will have only two terms $F_0\supset F_1$. We choose $F_1$ so that
$\bar{F}= F_0 / F_1$ is isomorphic to the field
algebra which is the starting point in \cite{SeTr}.
The right choice is $F_1$ generated by $\{\beta^\alpha$,
$\widehat{\gamma}_\alpha$, $\widehat{\psi}^{\alpha}$,
$\widehat{\psi}^{\und{\alpha}}$,
$\widehat{\psi}^{i}\}$. $\widehat{\gamma}_\alpha$ is constructed in
such a way that $F_1$ is a subcomplex of $F_0$:
\ben
\widehat{\gamma}_\beta= \gamma_\beta -\frac{1}{k} g_{\beta\und{\alpha}}
\left( {f^{\und{\alpha}\mu}}_{\bar{\epsilon}}\,
\widehat{\psi}^{\bar{\epsilon}}\,
c_\mu -(-)^{d_{\und{\alpha}}+d_\mu}\, {f^{\und{\alpha}\mu}}_\epsilon
\, \chi^\epsilon\, c_\mu \right)
\een
which implies $s(\widehat{\psi}^{\und{\alpha}})=
-k g^{\und{\alpha}\alpha}\, \widehat{\gamma}_\alpha$.
Indeed, with this definition of $F_1$, $\bar{F}$ is an affine superalgebra,
generated by
$\{\hat{\jmath}^a_\perp=\widehat{\jmath}^a
-\frac{1}{2k}{f^a}_{bc}\, \widehat{\psi}^c\widehat{\psi}^b$;
$b^\alpha$; $c_\alpha$; $\widehat{\psi}^i\}$.
This superalgebra is isomorphic to the one given in
\cite{SeTr} since
$\widehat{\jmath}^a_\perp$ commute with all the $\widehat{\psi}^a$ and
$dim(\Delta_0-\Pi_0)=dim\Delta_{1/2}$.

 On $\bar{F}$, the fields that belongs to $F_1$ can be set to zero.
This means for example that
$\gamma_\alpha\sim \frac{1}{k} g_{\alpha\bar{\beta}} \left(
{f^{\bar{\beta}\mu}}_{\bar{\imath}}\, \widehat{\psi}^{\bar{\imath}}
\, c_\mu -(-)^{d_{\bar{\beta}}+d_\mu}\, {f^{\bar{\beta}\mu}}_{\epsilon}
\, \chi^\epsilon\, c_\mu \right)$. Using these equivalences, one can
simplify $j_1$, so that on $\bar{F}$,
\ben
j_1(z)\sim (j^\alpha-\widehat{\chi}^\alpha)c_\alpha
-\frac{1}{2} (-)^{d_\gamma} {f^{\alpha\beta}}_\gamma\, b^\gamma\,
c_\beta c_\alpha -\frac{1}{k} g_{\alpha\und{\A}}
{f^{\und{\A}\B}}_{\bar{\imath}}\, \chi^\alpha\,
\widehat{\psi}^{\bar{\imath}}\, c_\B
\mb{with} \widehat{\chi}^\alpha= \frac{\nu}{k} {f^\alpha}_{\beta\gamma}
\, \chi^\beta\chi^\gamma
\een
One recognizes the form given in \cite{SeTr}. In particular, as
$\chi^\alpha=\delta_{\alpha,-}$, $\widehat{\chi}^\alpha$ is proportional to
$\delta_{\alpha,{--}}$.

Thus, $H^0(\bar{F},s)$ is the $W$ algebra that one gets from a quantum
Hamiltonian reduction of the superalgebra $\g$ with respect to the
$s\ell(2)$ subalgebra contained in the embedded $osp(1|2)$ algebra.

However, as $F_1$ is a subcomplex of $F_0$, we have the isomorphism:
$H^0(\bar{F},s)\simeq
\bar{H}$ where $\bar{H}=H^0(F_0,s)/H^0(F_1,s)$. We already know that
$H^0(F_0,s)$ is the supersymmetric $W$ algebra, and from
\ben
\begin{array}{ll}
s(\beta^\alpha)=\widehat{\psi}^\alpha-\chi^\alpha &
\Rightarrow\ \ \ s(\widehat{\psi}^\alpha)=0 \\
s(\psi^{\und{\alpha}})=-k g^{\und{\alpha}\alpha} \widehat{\gamma}_\alpha
&\Rightarrow\ \ \  s(\widehat{\gamma}_\alpha)=0 \\
s(\widehat{\psi}^i)=0 &
\end{array}
\een
it is clear that $H^0(F_1,s)$ is generated by the spin 1/2 fields
$\widehat{\psi}^i$ of the supersymmetric $W$ algebra\footnote{
The superfields $\hJ^\bA$ where $\bA \in \Pi_0$ are in $H^0(F_0,S)$, see
below}.

Thus, $\bar{H}$ is the super $W$ algebra after factorization of the
spin 1/2 fields. Because there is an isomorphism
between $\bar{H}$ and $H^0(\bar{F},s)$, we conclude that the
factorized super $W$ algebra (obtained from quantum $osp(1|2)$
reduction) and the $W$ superalgebra (that one gets from the quantum reduction
defined by $s\ell(2) \subset osp(1|2)$) are isomorphic.

\section{Explicit Construction of Super Generators.\label{s.4}}

\subsection{Spin 1 Super Generators.}

\indent

It is obvious from the tic-tac-toe construction,
that if the generator $t^\bA \in \Pi_0$, then the corresponding generator
in the $W$ algebra is just $W^\bA = \hJ_{hw}^\bA$; in other words, the
remaining
super Kac-Moody part of the algebra
is just the original affine current plus a ghost
term:  $W^\bA = J_{hw}^\bA + (-)^{d_\bA+d_\B}{f^{\bA\B}}_\G B^\G C_\B.$

Let us illustrate the tic-tac-toe construction by using it to find the
general form of a spin 1 generator in a supersymmetric $W$-algebra.
Let $\bA$ be an index corresponding to a highest-weight generator in a spin
$\hf$
representation $\r_{\hf}$. We wish to find $W^\bA = \sum_i (-)^i W^\bA_i$,
where $W^\bA_0 = \hJ_{hw}^\bA$ and
$S_1(W^\bA_0) = S_0(W^\bA_{1})$, $S_1(W^\bA_1) = 0$.
Before we show the result, let us give a few technical results that are used in
the calculations.

Using the supersymmetry and the invariance of the bilinear form $g^{ab}$,
as well as
the Jacobi identities, one can show that if $a$ and $b$ are indices for
generators with $osp(1|2)$ quantum numbers $(j,m)$ and $(j',m')$ respectively,
then
$g^{ab} \not = 0 \Rightarrow m+m'=0$ and $j=j'$.
{}For the $osp(1|2)$-generators, one
can even show that $g^{+a} \propto \D_{a,-}$ and $g^{++~a} \propto \D_{a,--}$.
We use this property to define a nonvanishing constant $\nu$ by
$g_{+a} = \nu \D_{a,-}$.
We define
furthermore\footnote{Using the Jacobi identities one can show that if
$J^a$ has $osp(1|2)$ quantum numbers $(j,m)$, then
$N_a = \frac{-1}{8\nu}\left ( (-)^{2(j-m)}(4j+1) - (4m+1) \right )$.}
$N_a = {f^{+~a}}_b {f^{-~b}}_a$ (no summation over $a$).

Another result that we will need is the relation between $(AB)_0(Z)$ and
$(BA)_0(Z)$. To find this relation we need the superspace version of the Taylor
expansion.
We write $Z = (z,\T)$, $W = (w,\E)$, $(Z-W) = z-w-\T\E$,
$D = \frac{\pa}{\pa \T} + \T \frac{\pa}{\pa z}$; it is not difficult to
show that
$$
F(Z) = \sum_{n=0}^\infty \frac{1}{n!} (Z-W)^n \pa^n
\left [ F(W) + (\T-\E)DF(W) \right ].
$$
Now, given two operators $A$ and $B$,
we can write the operator product expansion in the form
$$
A(Z)B(W) = \sum_{m=1}^\infty
\frac{ (AB)_m^0(W) + (\T-\E)(AB)_m^1(W)}{(Z-W)^m} + (AB)_0(W) + O(Z-W).
$$
The sum terminates at $m = h_A + h_B$ (or before), where $h_A$ and $h_B$ are
the conformal dimensions of the two operators.
Using the superspace Taylor expansion and the fact that
$A(Z)B(W) = \EP B(W)A(Z)$, $\EP = \pm 1$,
we find the following relation:
$$
(AB)_0 = \EP(BA)_0 + \EP \sum_{m=1}^\infty \frac{(-)^m}{m!} \pa^m
\left \{ (BA)_m^0 + (\T-\E)\left ( D(BA)_m^0 - (BA)_m^1 \right ) \right \}.
$$
Finally we need the superspace version of the Cauchy theorem. Defining
$$
\oint_W dZ = \oint_w \frac{dz}{2\pi i}\int d\T,
$$
one verifies that:
\bea
\oint_W dZ \frac{\T-\E}{(Z-W)^{(n+1)}} \Phi(Z)
& = & \frac{1}{n!} \pa^n \Phi(W), \nn
\oint_W dZ \frac{1}{(Z-W)^{(n+1)}} \Phi(Z)
& = & \frac{1}{n!} \pa^n D\Phi(W).
\eea

Using these results, we can now find $W^\bA = W^\bA_0 - W^\bA_1$,
$\bA \in \Delta_\hf$.
{}From eq. (\ref{brst-trans}) we have
\ben
S_1(\hJ^\bA) = - k g^{\bA\B}Dc_\B + {f^{\bA\B}}_\bG \hJ^\bG c_b
\label{ttt-1}
\een
$\hJ^\bA$ has bigrading $(1,-1)$; $S_1$ has bigrading (0,1) and $S_0$ has
bigrading (1,0), thus terms in $W^\bA_1$ must have bigrading (0,0).
We find that
\bea
S_0({f^{-~\bA}}_\bB D\hJ^\bB) & = & \hf g^{\bA\B} Dc_\B, \nn
\frac{1}{\nu N_{\bB}-m_\bB} S_0( {f_{\bG\bB}}^\bA {f^{-~\bB}}_{\bE}
\hJ^\bE \hJ^\bG)
& = & -2 \nu {f^{\bA\B}}_\bG \hJ^\bG c_\B.
\label{Wa1}
\eea
where $\bB \in \Delta_{m_\bB}$.
Note that if $t^\bB \in \r_{j_\bB}$ then
${f^{-~\bB}}_{\bE} \not = 0 \Rightarrow j_\bB,m_\bB > 0
\Rightarrow \nu N_\bB - m_\bB \not = 0$.
Comparing with eq. (\ref{ttt-1}) we see that
$$
W^\bA_1 = -2k {f^{-~\bA}}_\bB D\hJ^\bB
- \frac{1}{2(\nu N_\bB - m_\bB)} {f_{\bG\bB}}^\bA {f^{-~\bB}}_{\bE}
\hJ^\bE \hJ^\bG.
$$
$S_1(W^\bA_1)$ vanishes, so
we can now write the general form of a spin $1$ generator in a supersymmetric
$W$-algebra as
\ben
W^\bA = \hJ_{hw}^\bA + 2k {f^{-~\bA}}_\bB D\hJ^\bB
+ \frac{1}{2(\nu N_\bB - m_\bB)} {f_{\bG\bB}}^\bA {f^{-~\bB}}_{e} \hJ^e
\hJ^\bG.
\quad \quad {\rm for}~\bA\in\Delta_\hf.
\label{Wa1/2}
\een
Actually the calculation for $W^\bA_1$ did not depend on the fact that the
grade
of $\bA$ was 1/2, except for some normalizations. If $\bA$ is the index of a
highest weight generator in a representation $\r_j$, $j \not = 0$,
then the first equation in (\ref{Wa1}) is replaced by
\bea
S_0({f^{-~\bA}}_\bB D\hJ^\bB) & = & j g^{\bA\B} Dc_\B.
\eea
We see that for general $\bA$, where $t^\bA$ is a highest weight generator,
the expression for $W^\bA$ starts with
\bea
W^\bA = \hJ_{hw}^\bA
+ \frac{k}{j} {f^{-~\bA}}_\bB D\hJ^\bB
+ \frac{1}{2(\nu N_\bB - m_\bB)} {f_{\bG\bB}}^\bA {f^{-~\bB}}_{e} \hJ^e \hJ^\bG
+ \cdots. \nn
t^\bA \in \r_j, \quad \quad j \not = 0
\eea

\subsection{Energy-Momentum Tensor and Quantum Miura Transformation.}

\indent

In principle we could find the complete
expression for a general generator in any supersymmetric $W$ algebra
obtainable by hamiltonian reduction, but the calculations quickly becomes
intractable.
However, in the special case of the energy-momentum tensor we can
find the complete expression $W^{++} = W^{++}_0 - W^{++}_1 + W^{++}_2$:
($\mu \equiv {f^{+~+}}_{++}$, $\mu \not = 0$)
\bea
W^{++}_0 & = & \hJ^{++}, \nn
W^{++}_1 & = & \frac{k}{\nu\mu} D\hJ^+
- \frac{1}{\nu\mu} {f^+}_{\bA\bB} \hJ^\bB\hJ^\bA, \nn
W^{++}_2 & = & \frac{1}{\nu^2\mu} \left (
2k(k+\Si) \partial \hJ^0 - k g_{\bB_0\bA_0} \hJ^{\bA_0}D\hJ^{\bB_0}
+ \frac{1}{3} f_{\bG_0\bB_0\bA_0} \hJ^{\bA_0} \hJ^{\bB_0}  \hJ^{\bG_0} \right
),
\eea
$\Si$ is a contribution from the normal ordering,
and can be expressed as
$$
\Si = \nu\sum_{\bA\in\Delta_0} (-)^{d_\bA}N_\bA.
$$
After
normalization we find:
\bea
T & = & \frac{\nu^2\mu}{2k^2} \hJ^{++} - \frac{\nu}{2k} D\hJ^+
+ \frac{\nu}{2k^2} {f^+}_{\bA\bB} \hJ^\bB\hJ^\bA \nn
& & + \frac{k+\Si}{k}\pa \hJ^0 - \frac{1}{2k} g_{\bB_0\bA_0}
\hJ^{\bA_0}D\hJ^{\bB_0}
+ \frac{1}{6k^2} f_{\bG_0\bB_0\bA_0} \hJ^{\bA_0} \hJ^{\bB_0}  \hJ^{\bG_0}.
\label{T}
\eea

As mentioned above, it is possible to show that all the generators of
the $W$-algebra includes a term of zero grade. It is easy to see, that if we
restrict to
these zero-grade terms, we have a realization of the $W$-algebra.
Thus, if we write the grade zero part of the affine Lie superalgebra in
terms of free fields using the Wakimoto construction\footnote{In \cite{It}
the Wakimoto construction was done for Lie superalgebras
in the non-supersymmetric case. To get a supersymmetric version, it
suffices to add spin 1/2 fermions and bosons in the adjoint representation, see
\cite{DiKnPeRo,Fu,GoOlWa}. We thank K. Ito for drawing our attention to this.},
we immediately get a free
field realization of the $W$ algebra. This construction is called the quantum
Miura transformation. In the energy momentum tensor above, the grade zero part
was
\ben
T_0 = \frac{k+\Si}{k}\pa \hJ^0 - \frac{1}{2k} g_{\bB_0\bA_0}
\hJ^{\bA_0}D\hJ^{\bB_0}
+ \frac{1}{6k^2} f_{\bG_0\bB_0\bA_0} \hJ^{\bA_0} \hJ^{\bB_0}  \hJ^{\bG_0}.
\label{T0}
\een
This should be compared with the classical Miura transformation of the energy
momentum tensor (written in terms of the affine currents in the grade zero
part) \cite{DeRaSo}:
\ben
T_{class} = \pa J^0 - \frac{1}{2k} g_{\bB_0\bA_0} J^{\bA_0} DJ^{\bB_0}
+ \frac{1}{6k^2} f_{\bG_0\bB_0\bA_0} J^{\bA_0} J^{\bB_0}  J^{\bG_0}.
\een
{}From equation (\ref{T0}) we can easily calculate the central charge of the
algebra :
\bea
\hat{c} & = & \mbox{sdim }\g_0 + \frac{2}{3k}f^{\bA_0\bB_0\bG_0}
f_{\bG_0\bB_0\bA_0}
- 8 g^{00} (k + 2 \Si + \frac{\Si^2}{k}).
\label{cc}
\eea
$\g_0$ denotes the grade zero part of the superalgebra $\g$, and sdim is the
superdimension. We know from the normalization of the embedded $osp(1|2)$
algebra that $g^{00} = \pm \hf$.

Note that for those algebras which in addition to the spin 3/2
super energy momentum tensor contains only a number of spin one and spin 1/2
generators, the
superconformal, quasi-superconformal and $\Z_2 \times \Z_2$
superconformal algebras,
we have found the general expressions for all
generators in the algebras, and using the quantum Miura transformation it is
easy to write down the corresponding free field realizations.

\section{Example: $B(n|1)$.\label{s.5}}

\indent

As an example we will take the hamiltonian reduction of $B(n|1) = osp(2n+1|2)$.
As $osp(1|2)$--subalgebra we choose the regular embedding of $osp(1|2)$ into
$osp(2n+1|2)$, leading to a super Kac-Moody extended superconformal algebra
\cite{FrRaSo}. We parametrize the affine current as follows:

\ben
\label{Jbn1}
\!\!\!\!\!
J(Z) = \left ( \begin{array}{ccccccc|cc}
J^{11}   & J^{12} & \cdots & J^{1,n+1} & \cdots & J^{1,2n} & 0 &
J^{1-}  & J^{1+} \\
J^{21}   & J^{22} & \cdots & J^{2,n+1} & \cdots & 0 & -J^{1,2n} &
J^{2-}  & J^{2+} \\
\vdots     & \vdots   &        & \vdots &            & \vdots      &
\vdots  & \vdots & \vdots\\
J^{n+1,1} & J^{n+1,2} & & 0 & & -J^{2,n+1} & -J^{1,n+1} &
J^{(n+1)-} & J^{(n+1)+} \\
\vdots     & \vdots   &        & \vdots &            & \vdots      &
\vdots  & \vdots &\vdots \\
J^{2n,1} &        0 &  & -J^{n+1,2} & \cdots &  -J^{22} & -J^{12} &
J^{2n-} & J^{2n+} \\
0 & -J^{2n,1} & \cdots & -J^{n+1,1} & \cdots & -J^{21} & -J^{11} &
J^{(2n+1)-} & J^{(2n+1)+} \\
\hline
-J^{1+} & -J^{2+} & \cdots & -J^{(n+1)+} & \cdots & -J^{2n+} & -J^{(2n+1)+} &
J^0 & J^{++} \\
J^{1-} & J^{2-} & \cdots & J^{(n+1)-} & \cdots & J^{2n-} & J^{(2n+1)-} &
J^{--} & -J^0
\end{array} \right )
\een

\noindent
The fermionic roots of $osp(1|2)$ are $J^\pm = J^{(n+1)\pm}$.
{}For $i+j > 2n+2$, we define $J^{ij} = - J^{2n+2-i,2n+2-j}$.

The algebra splits into
representations of $osp(1|2)$: a spin 1 representation $\r_1$ which is just
the embedded $osp(1|2)$ itself,
$\r_1 = \{J^{++},J^{(n+1)+},J^0,J^{(n+1)-},J^{--} \}$, $2n$ spin $\hf$
representations each of the form
$\r_\hf = \{J^{i+},J^{(n+1)i},J^{i-}\},~i \not = n+1$, and $n(2n-1)$
singlet representations $\r_0 = \{ J^{ij} \},~i,j \not = n+1,~i+j<2n+2$.

We find that with this parametrization, the operators product expansions are
($Z_{12}\equiv Z_1-Z_2$, $\T_{12}\equiv \T_1-\T_2$, operators on the right hand
side are taken in the $Z_2$ variable):
\bea
J^0(Z_1) J^0(Z_2)         & = & \frac{-k/2}{Z_{12}}             +\cdots\nn
J^0(Z_1) J^{++/--}(Z_2)   & = & \toz (\pm J^{++/--})            +\cdots\nn
J^{++}(Z_1) J^{--}(Z_2)   & = & \frac{-k}{Z_{12}} + \toz 2J^0   +\cdots\nn
J^{ij}(Z_1) J^{kl}(Z_2)    & = & \frac{\frac{k}{2} (\D_{il}\D_{jk} -
                      \D_{j,2n+2-l}\D_{i,2n+2-k})}{Z_{12}}             \nn
&&         \hspace{-10mm} + \toz \hf \left (
            \D_{il}J^{kj} - \D_{kj}J^{il}
              - \D_{i,2n+2-k} J^{2n+2-l,j}
      + \D_{j,2n+2-l} J^{i,2n+2-k} \right )                     +\cdots\nn
J^0(Z_1) J^{i\pm}(Z_2)    & = & \toz (\pm \hf J^{i\pm})         +\cdots\nn
J^{++}(Z_1) J^{i-}(Z_2)   & = & \toz (-J^{i+})           +\cdots\nn
J^{--}(Z_1) J^{i+}(Z_2)   & = & \toz (-J^{i-})           +\cdots\nn
J^{ij}(Z_1) J^{k-}(Z_2) & = & \toz
    \hf ( \D_{i,2n+2-k} J^{(2n+2-j)-} - \D_{jk} J^{i-} )        +\cdots\nn
J^{ij}(Z_1) J^{k+}(Z_2) & = & \toz
   \hf ( \D_{i,2n+2-k} J^{(2n+2-j)+} - \D_{jk} J^{i+} )        +\cdots\nn
J^{i+}(Z_1) J^{j-}(Z_2)   & = & \frac{-\frac{k}{2}\D_{i,2n+2-j}}{Z_{12}}
                           +\toz \hf(J^{2n+2-i,j}-\D_{i,2n+2-j} J^0) +\cdots\nn
J^{i\pm}(Z_1) J^{j\pm}(Z_2) & = & \toz (\mp \hf \D_{i,2n+2-j}
J^{++/--})                                                      +\cdots.
\eea
The constraints that we wish to impose are $J^{(n+1)-}(Z) = 1$  and
$J^{--}(Z) = J^{i-}(Z) = 0,~i\not = n+1$.
We introduce ghost super--fields $(B^{i-},C_{i-})$ and
$B^{--},C_{--}$; $C_{i-}$ and $B^{--}$ are bosonic while $C_{--}$
and $B^{i-}$ are fermionic. The BRST currents are
\ben
J_0(Z) = - C_{(n+1)-}(Z)
\een
and
\ben
J_1(Z) = J^{i-}(Z)C_{i-}(Z) + J^{--}(Z)C_{--}(Z)
+ \frac{1}{4}~B^{--}C_{i-}C_{(2n+2-i)-}(Z).
\een

The``hatted" currents are
\bea
\hJ^{++}  & = & J^{++},                   \nn
\hJ^{i+}   & = & J^{i+} + B^{i-}C_{--},       \nn
\hJ^{0} & = & J^0 + \hf (B^{i-}C_{i-})_0 - (B^{--}C_{--})_0,  \nn
\hJ^{ij} & = & J^{ij} + \hf B^{i-}C_{j-} - \hf B^{(2n+2-j)-}C_{(2n+2-i)-}
\nn
\hJ^{i-}   & = & J^{i-} + \hf B^{--}C_{(2n+2-i)-},     \nn
\hJ^{--}  & = & J^{--}.
\eea

The $W$-algebra generators corresponding to the singlet representations of the
$osp(1|2)$ are just the hatted currents :
\ben
W^{ij} = \hJ^{ij} ~=~
J^{ij} + \hf B^{i-}C_{j-} - \hf B^{(2n+2-j)-}C_{(2n+2-i)-},~i,j \not = n+1.
\een
They form a supersymmetric $so(2n)$--subalgebra of the resulting $W$--algebra.

The form of
the $W$--generators corresponding to the highest weights of the spin $\hf$
representations was given in eq. (\ref{Wa1/2}), and can be written as
\ben
W^{i+} = \hJ^{i+} - k D\hJ^{i,n+1} - (\hJ^{i,k}\hJ^{k,n+1})_0
- (\hJ^0\hJ^{i,n+1})_0
\een

Finally we have the energy momentum tensor, which after normalizations is given
in eq. (\ref{T}); we find
\bea
T & = & - \frac{1}{k^2} J^{++}
+ \frac{1}{k} D\hJ^{(n+1)+} + \frac{2}{k^2} \hJ^{(n+1,j}\hJ^{j+}
+ \frac{2}{k^2} \hJ^0 \hJ^{(n+1)+} \nn
& & + \frac{k-n+1/2}{k} \pa \hJ^0
- \frac{1}{2k} \tr (\hM D\hM)
+ \frac{1}{k} (\hJ^0 D\hJ^0)_0
- \frac{1}{3k^2} \tr (\hM\hM\hM),
\eea
where $\hM$ is the ``hatted" submatrix corresponding to the $so(2n+1)$-part in
(\ref{Jbn1}).
The central charge is found to be
$\hat{c} = 4k - 8n + 5 + \frac{1}{k}(n-\hf)^2$. A free field realization of the
algebra can be found by inserting a free field realization of the grade zero
part into the expressions above; we can take a Wakimoto realization
of the super-symmetric $so(2n+1)^{(1)}$, and realize $J^0$ as a free super
field $J^0 \rightarrow \sqrt{\frac{k}{2}} D\Phi$.

\section{Conclusion.\label{s.6}}

In this paper, we have considered the quantum hamiltonian reduction of
supersymmetric affine Lie superalgebras, giving as a result the quantized
supersymmetric $W$ algebras.
We have quantized all the manifestly supersymmetric $W$ algebras which can be
obtained by Hamiltonian reduction of affine Lie superalgebras.
We have shown that using a suitable cohomological procedure, we can reduce the
supersymmetric quantum hamiltonian reduction defined by an $osp(1|2)$ embedding
into a Lie superalgebra $\g$, to the non-supersymmetric hamiltonian reduction
defined by the embedding of $s\ell(2) \subset osp(1|2)$ into $\g$.
We have given
an explicit expression for the energy momentum tensor and the central charge
of all supersymmetric $W$ algebras, as well as
general expressions for all generators in the
superconformal, quasi-superconformal and $\Z_2 \times \Z_2$ superconformal
algebras. We have described the quantum Miura transformation, and described the
example of the hamiltonian reduction of $osp(2n+1|2)^{(1)}$ in some detail.

\newpage
%\vspace{5mm}
{\bf Acknowledgements}\\[2pt]
The authors wish to acknowledge helpful discussions with F. Delduc, and
to thank Kris Thielemans for supplying us with his Mathematica package for
calculating super OPE's.
JOM wishes to thank the \LAP, where this work was done, for its kind
hospitality, and the Danish Research Academy for its financial support.

\indent


{\bf References}

\begin{thebibliography}{99}
\bibitem{BeOo} M. Bershadsky and H. Ooguri. \CMP{126}, (1989) 49.
\bibitem{BaTjVDr} F.A. Bais, T. Tjin and P. van Driel, \NPB{B357} (1991) 632.
\bibitem{BaFeFoRaWi} L. Feh\'er, L. O'Raiffeartaigh, P. Ruelle
and I. Tsutsui, {\it Phys. Rep.} {\bf 222} (1992) 1, and references therein.
\bibitem{BoSc} P. Bouwknegt and K. Schoutens, {\it Phys. Rep.} {\bf 223}
(1993) 183.
\bibitem{FeFr} B.L. Feigin and E. Frenkel, \PLB{246}, (1990) 75.
\bibitem{BoMcPi} P. Bouwknegt, J. McCarthy and K. Pilch, in proceedings
{\it Strings and Symmetries}, N. Berkovitz et al. (eds.), World Scientific
1991.
\bibitem{FrKaWa} E. Frenkel, V. Kac and M. Wakimoto, \CMP{147} (1992) 295.
\bibitem{BoTj} J. de Boer and T. Tjin, ``The Relation Between Quantum W
Algebras
and Lie Algebras", {\it preprint} THU-93/05;
\CMP{158} (1993) 485.
\bibitem{SeThTr} A. Sevrin, K. Thielemans and W. Troost, \NPB{407}, (1993) 459.
\bibitem{SeTr} A. Sevrin and W. Troost, \PLB{315}, (1993) 304.
\bibitem{DeRaSo} F. Delduc, E. Ragoucy and P. Sorba, \CMP{146}, (1992) 403.
\bibitem{FrRaSo} L. Frappat, E. Ragoucy and P. Sorba, \CMP{157}, (1993) 499.
\bibitem{DiKnPeRo}P. Di Vecchia, V.G. Knizhnik, J.L. Petersen and P. Rossi,
\NPB{253}, (1985) 701.
\bibitem{FrLi} E.S. Fradkin and V.Ya. Linetskii, \PLB{291} (1992) 71.
\bibitem{ItMaPe} K. Ito, J.O. Madsen and J.L. Petersen,
{\it String Theory, Quantum Gravity and the Unification of the Fundamental
Interactions},  M. Bianchi, F. Fucito, E. Marinari and A. Sagnotti (eds.),
World Scientific 1993, p. 302;\\
K. Ito, J.O. Madsen and J.L. Petersen, \PLB{318} (1993) 315.
\bibitem{KoSt}B. Kostant and S. Sternberg, {\it Ann. Phys.} {\bf 176} (1987)
49.
\bibitem{Ra} E. Ragoucy, \NPB{411}, (1994) 778.
\bibitem{DeTh}A. Deckmyn and K. Thielemans, ``Factoring out Free Fields", {\it
preprint} KUL-TH-93-26.
\bibitem{It} K. Ito, \IJMPA{7}, (1992) 4885.
\bibitem{Fu} J. Fuchs, \NPB{318}, (1989) 631.
\bibitem{GoOlWa} Goddard, Olive and Waterson \CMP{112}, (1987) 591.
\end{thebibliography}
\end{document}